\documentclass[aps,prl,tighten,superscriptaddress,notitlepage]{revtex4-1}
\usepackage{amsmath,fullpage,graphicx,amssymb,bm}
\usepackage{fancyvrb,setspace,titlesec}
\usepackage{hyperref}

\titleformat{\section}[runin]{\normalfont\bfseries}{\thesection}{0.5em}{}[.]
\usepackage{aas_macros}
\begin{document}
\date{\today}

\title{\vspace{-1.1cm} \Large \textbf{\textsc{Prospects for Indirect
      Detection of Dark Matter with CTA} \vspace{-0.3cm}}}

\author{M. Wood} 
\email{mdwood@slac.stanford.edu}
\affiliation{KIPAC/SLAC National Accelerator Laboratory, Menlo Park, CA 94025, USA}

\author{J. Buckley} 
\affiliation{Department of Physics, Washington University, St. Louis,
  MO, 63130, USA }

\author{S. Digel} 
\affiliation{KIPAC/SLAC National Accelerator Laboratory, Menlo Park,
  CA 94025, USA}

\author{S. Funk} 
\affiliation{KIPAC/SLAC National Accelerator Laboratory, Menlo Park,
  CA 94025, USA}

\author{D. Nieto}
\affiliation{Physics Department, Columbia University, New York, NY 10027, USA}

\author{M. A. S\'{a}nchez-Conde} 
\affiliation{KIPAC/SLAC National Accelerator Laboratory, Menlo Park, CA 94025, USA}

\begin{abstract}
  We discuss the prospects for indirect detection of dark matter (DM)
  with the Cherenkov Telescope Array (CTA), a future ground-based
  gamma-ray observatory that will be sensitive to gamma rays in the
  energy range from a few tens of GeV to 100 TeV.  We consider the
  detectability of DM annihilation in different astrophysical targets
  with a focus on the Galactic Center (GC) region.  With a deep
  observation of the GC, CTA will be sensitive DM particles with mass
  greater than $100$~GeV and an annihilation cross section close to
  the thermal relic value.
\end{abstract}

\maketitle

\vspace{-1.1cm} 

\section{Introduction}

Strong evidence indicates that most of the matter in the Universe is
dark. Indeed, in the standard cosmology, $\sim$27\% of the Universe
consists of non-baryonic dark matter (DM)
\citep{2013arXiv1303.5076P}. DM has not been conclusively detected in the
laboratory yet, but its gravitational effects have been observed on
spatial scales ranging from the inner kiloparsecs of galaxies out to
Mpc and cosmological scales. Also, the only way to explain the
formation of large scale structure is by requiring that the dominant
component of matter in the Universe is cold DM.
% One of the first steps in the DM paradigm was taken by F. Zwicky in
% the 1930s to explain the velocity dispersion in galaxy
% clusters. Today, the most conclusive observations in this sense come
% from the rotational speeds of galaxies, the orbital velocities of
% galaxies within clusters, gravitational lensing, satellite
% kinematics, the cosmic microwave background and large scale
% structure.  However, we still do not know what the DM is made
% of. Well-motivated potential physics beyond the Standard Model of
% particle physics can provide non-baryonic DM candidates that fulfill
% all the cosmological requirements.  A plethora of possible DM
% candidates have already been proposed.
Observations of separate distributions of the baryonic and
gravitational mass in galaxy clusters indicate that the DM is likely
composed of particles with a low interaction cross section relative to
ordinary matter.  Particle physics theory predicts degrees of freedom
for new particles at the 100 GeV to 10 TeV scale to solve the hierarchy
problem in the Standard Model \citep{2013arXiv1302.6587F}.
Remarkably, weakly interacting 100~GeV-scale particles would naturally
result in the correct relic abundance.  The concordance of the diverse
astrophysical data, together with compelling theoretical arguments
provide a strong case for DM searches aimed at the detection of a
thermal relic with weak-scale interactions with ordinary matter.

One of the most popular candidates for DM is the class of models known
as weakly interacting massive particles (WIMPs).  In regions of high
DM density the annihilation (or decay) of WIMPs into Standard Model
particles could produce a distinctive signature in gamma rays potentially
detectable with ground- and space-based gamma-ray observatories. 
%covering the energy range from 100 MeV to 100 TeV.
In fact, almost any annihilation channel will eventually produce
gamma-rays either through pion production (for hadronic channels), or
final state bremmstrahlung and inverse Compton from leptonic channels.
Moreover, the spectrum from annihilation would be universal, with the
same distinctive shape detected in every DM halo.  The measurement of
the gamma-ray signature would also complement direct searches by
providing a strong constraint on the WIMP mass.  A detection with both
techniques would uniquely reveal both the mass and scattering cross
section of the WIMP particle.

% Unlike the signals that can be measured by direct DM detection
% experiments, the gamma-ray signature would provide strong
% constraints on the particle mass and help to identify the particle
% through the different kinematic signatures predicted for different
% annihilation channels.

The planned Cherenkov Telescope Array (CTA)
\citep{2011ExA....32..193A} is designed to have sensitivity over the
energy range from a few tens of GeV to 100 TeV.  To achieve the best
sensitivity over this wide energy range CTA will include three
telescope types: Large Size Telescope (LST, 23 m diameter), Medium
Size Telescope (MST, 10-12 m) and Small Size Telescope (SST, 4-6 m).
Over this energy range the point-source sensitivity of CTA will be at
least one order of magnitude better than current generation imaging
atmospheric Cherenkov telescopes such as H.E.S.S., MAGIC, and VERITAS.
CTA will also have an angular resolution at least 2--3 times better
than current ground-based instruments, improving with energy from
0.1$^\circ$ at 100 GeV to better than 0.03$^\circ$ at energies above 1
TeV.

\section{Targets for Indirect DM Searches}

The gamma-ray flux from DM annihilations scales with the integral of
the square of the DM density along the line of sight to the source
($J$). Thus, the detectability of the DM signal from a given target
depends critically on its DM distribution.  The ideal targets for DM
annihilation searches are those that have both a large value of $J$
and relatively low astrophysical gamma-ray foregrounds.  These
criteria have motivated a number of Galactic and extragalactic targets
including the Galactic Center (GC), dwarf spheroidal satellite
galaxies of the Milky Way (dSphs), and galaxy clusters.  While the
sensitivity of the signal to the DM halo profile is a source of
significant systematic uncertainty, it also provides an avenue for
inferring the DM halo profile from the shape of the gamma-ray
emission.  The detection of a distinctive spatial morphology would
definitively connect the detected particle to the missing
gravitational mass in galaxies.

The GC is expected to be the brightest source of DM annihilations in
the gamma-ray sky by at least two orders of magnitude. Although the
presence of many astrophysical sources of gamma-ray emission toward
the inner Galaxy make disentangling the DM signal difficult in the
crowded GC region, the DM-induced gamma-ray emission is expected to be
so bright there that one can realize strong upper limits at the level
of the natural cross section $\langle \sigma v\rangle \sim
10^{-26}{\rm cm}^{3}{\rm s}^{-1}$.  In addition, with the improved
angular resolution of CTA, the astrophysical foregrounds can be more
easily identified and separated from the diffuse annihilation signal.
% several astrophysical processes being at work in the crowded GC
% region might make extremely difficult to disentangle the DM signal
% from conventional emissions, the DM-induced gamma-ray emission is
% expected to be so large there that the search could be still
% worthwhile.
Also, the large concentration of baryons in the innermost region of
the Galaxy might act to further increase the expected DM annihilation
flux by making the inner slope of the DM density profile steeper
\citep{2004ApJ...616...16G}.  While the exact role of baryons is not
yet well understood, new state-of-the art numerical studies of
structure formation that include baryonic physics along with the
non-interacting DM are beginning to provide valuable insights
\citep{2011MNRAS.416.1118C,2012ApJ...744L...9M}.

N-body simulations of galactic structure formation show the evolution
of the cold DM distribution from an initial state of almost
homogeneous density into a present epoch of hierarchically assembled
clustered state embedded into a main smooth galactic
halo~\citep{2008Natur.454..735D}. The mass range of this wealth of
subhalos spans all resolved mass scales. In this context, dSphs are
interpreted as large DM subhalos of the Milky Way. dSphs are
attractive for DM searches in gamma rays due to their close proximity,
high DM content, and the absence of intrinsic sources of gamma-ray
emission.  Because they are highly DM-dominated, the DM mass on small
spatial scales ($\sim$100 pc) can be directly inferred from
measurements of their stellar velocity dispersions.  The uncertainty
of the line of sight distribution of DM for these systems is therefore
much less than for other candidates. Additionally, smaller DM subhalos
may not have attracted enough baryonic matter to ignite star-formation
and would therefore be invisible to most astronomical observations
from radio to X-rays. All-sky monitoring instruments sensitive at
gamma-ray energies, like Fermi-LAT, may detect the DM annihilation
flux from such subhalos~\citep{2010PhRvD..81d3532K}, while follow-up
observations with CTA would characterize the distinctive spectral
cut-off that would eventually determine the DM particle mass (see
\citep{2013arXiv1305.0312N} for further discussion of this strategy).

Galaxy clusters are another potential class of extragalactic targets
for DM searches.  The best candidates are nearby galaxy clusters such
as Virgo, Fornax, Perseus, and Coma.  Evaluated on the basis of the
smooth DM component, the annihilation signals of the best galaxy
cluster candidates are fainter on average than for dSphs.  Yet, when
the contributions from DM subhalos are included the expected DM
signals from these systems could be significantly greater
\citep{2011JCAP...12..011S}.  In contrast, this substructure boost is
expected to be only a small effect for dSphs and the GC.

\section{Current Constraints on the DM Annihilation Cross Section}

\begin{figure*}[!t]
  \centering
  \includegraphics[width=0.99\textwidth]{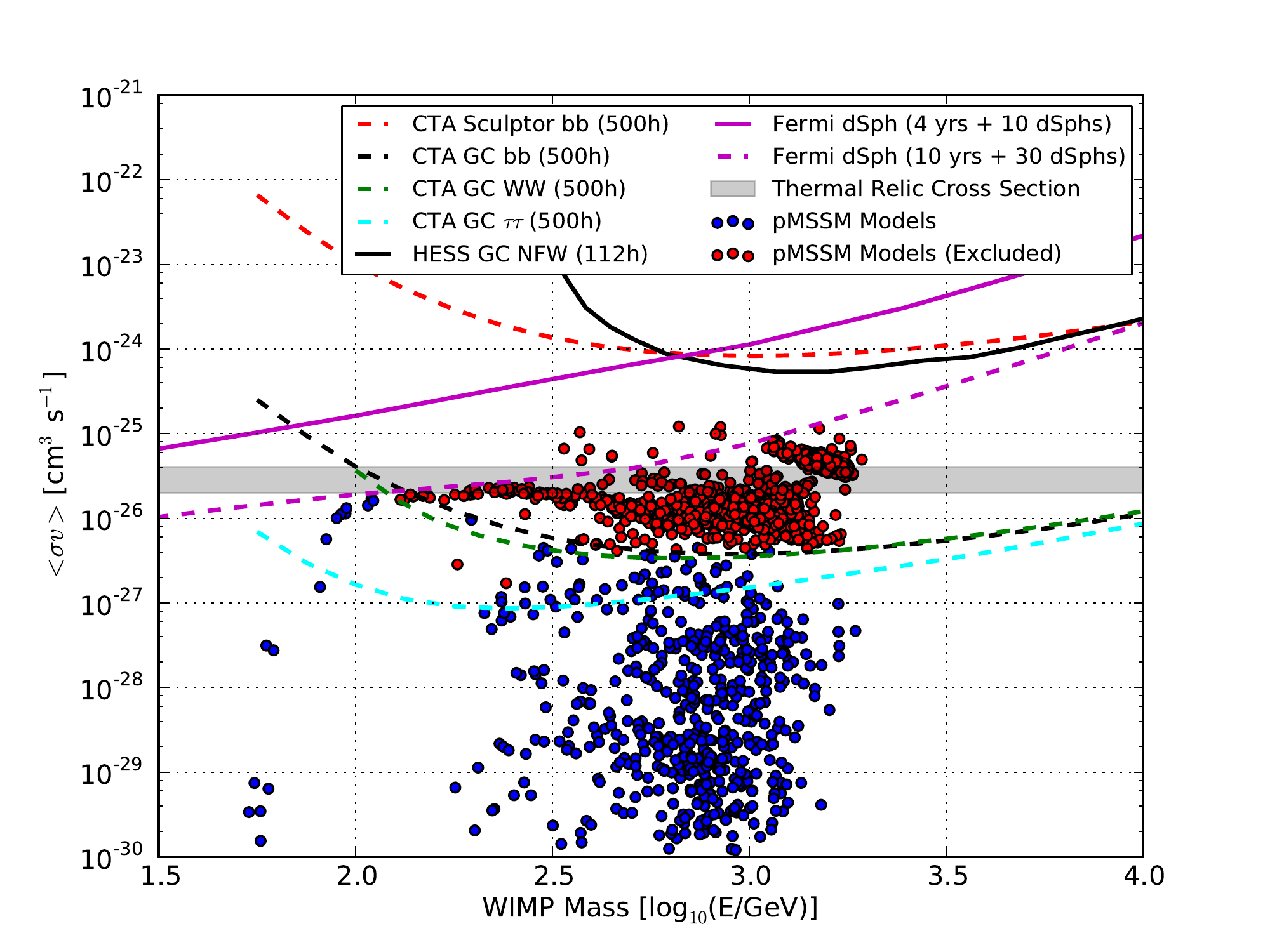}
  \caption {\label{FIG:DMLIMITS} Comparison of current (solid lines)
    and projected (dashed lines) limits on the DM annihilation cross
    section from different gamma-ray searches as a function of WIMP
    mass.  Limits for Fermi (magenta lines) and H.E.S.S. (solid black
    line) are calculated for a 100\% branching ratio to $bb$.
    Projected limits for CTA are shown for WIMP annihilation to $bb$
    and a 500 hour observation of Sculptor (red dashed line) and for
    WIMP annihilation to $bb$ (black dashed line), $W^{+}W^{-}$ (green
    dashed line), and $\tau^{+}\tau^{-}$ (cyan dashed line) and a 500
    hour observation of the GC.  The calculation of the annihilation
    flux for the GC region assumes an NFW MW halo profile with a scale
    radius of 20 kpc and DM density at the solar radius of 0.4
    GeV~cm$^{-3}$.  Filled circles represent pMSSM models satisfying
    WMAP7 constraints on the relic DM density and experimental
    constraints from ATLAS and CMS SUSY searches and XENON100 limits
    on the spin-independent WIMP-nucleon cross section
    \citep{2011EPJC...71.1697C,2012EPJC...72.2156C}.  Models indicated
    in red would be excluded by the CTA 95\% C.L. upper limit from a
    500 hour observation of the Galactic Center.}
\end{figure*}

Searches for the DM gamma-ray annihilation signature have been
conducted by all current ground-based Cherenkov telescope
observatories, H.E.S.S., MAGIC, and VERITAS.  VERITAS and MAGIC have
conducted several observation campaigns on northern hemisphere dSphs
including Coma Berenices, Willman I, Draco, Ursa Minor, and Segue 1
\citep{2009ApJ...697.1299A,2010ApJ...720.1174A,2011JCAP...06..035A,2012PhRvD..85f2001A}
while H.E.S.S. has conducted observations of the southern hemisphere
dSphs Sagittarius, Canis Major, Sculptor, and Carina
\citep{2008APh....29...55A,2009ApJ...691..175A,2011APh....34..608H}.
% For the less massive dSphs such as Segue 1 and Coma Berenices, the
% relatively small number of stars that can be used to constrain the
% gravitional potential introduces significant uncertainty on the
% inferred limit on the annihilation cross section.  The VERITAS
% collaboration has published the most constraining limits on the WIMP
% annihilation cross section using observations of Segue 1 which
% require $\sigma v < 10^{-23}$ cm$^2$ s$^{-1}$ for a WIMP mass of
% 1~TeV.
A search for a DM signal in the annular region of 0.3$^\circ$--1.0$^\circ$
around the GC conducted using 112~h of H.E.S.S. observations
\citep{2011PhRvL.106p1301A} currently sets the most
constraining limits on the DM annihilation cross section for WIMP masses
above 1~TeV, reaching $\sim 7 \times 10^{-25}$ cm$^2$ s$^{-1}$ at 1 TeV
for WIMPs annihilating through the $bb$ channel (Fig.~\ref{FIG:DMLIMITS}).

The data from the Large Area Telescope (LAT) on the {\it Fermi}
satellite have been used to search for DM annihilations in the energy
range 100 MeV -- 100 GeV by looking for signatures of point-like
emission in dSphs \citep{2011PhRvL.107x1302A} and galaxy clusters as
well as diffuse gamma-ray emission from the Galactic DM halo
\citep{2012PhRvD..86b2002A,2012ApJ...761...91A} and from DM
annihilation at cosmological distances \citep{2010JCAP...04..014A}.
% Because of its wide-field of view, survey mode operation, and low
% particle background contamination, the Fermi-LAT is well suited to
% searching for point-like DM sources as well as diffuse emission from
% DM annihilation in the Milky Way halo.  
The upper limits on the annihilation cross section derived from the
analysis of dSphs are among the most constraining for WIMP models with
masses below 300 GeV.  As compared with searches for DM signatures in
the Milky Way halo, these limits also have smaller systematic
uncertainties associated with modeling astrophysical foregrounds and
the DM distribution in the target region.  Figure \ref{FIG:DMLIMITS}
shows upper limits on the annihilation cross section derived from a
combined analysis of 10 dSphs and 4 years of LAT data. These are among
the most stringent limits on the DM annihilation cross section
obtained so far by any technique (including the LHC, or direct
detection experiments). Also shown in the figure is the projected LAT
limit with 10 years of data and an additional 20 dSphs which could be
discovered in future optical surveys.

\section{Projected Sensitivity of CTA for DM Searches}

% (commented out the text below because it was said already in the Introduction}
%CTA will have significantly greater sensitivity to DM than existing
%ground-based instruments with a ten times greater point-source
%sensitivity and a significantly lower energy threshold of 30--50~GeV.

%The potential of CTA for DM searches and testing other exotic physics
%has been studied in detail by \citet{2012arXiv1208.5356D} using the
%projected performance of the array E and I configurations
%\citep{2012arXiv1210.3503B} with 18--25 MSTs
%and 3--4 LSTs.  For the study presented here we have

The potential of CTA for DM searches and testing other exotic physics
has been studied in detail by \citep{2013APh....43..189D} using the
projected performance of several alternative array configurations
\citep{2012arXiv1210.3503B} with 18--37 MSTs and different
combinations of SSTs and LSTs. For the study presented here we have
considered the performance of a candidate CTA configuration with 61
MSTs corresponding to the baseline MST array with an additional US
contribution of 36 MSTs \citep{2012AIPC.1505..765J}.  This
configuration has comparable point-source sensitivity to previously
studied CTA configurations below 100~GeV but {\bf 2--3 times better
  point-source sensitivity between 100~GeV and 1~TeV.}

Figure \ref{FIG:DMLIMITS} shows the projected sensitivity of our
candidate CTA configuration to a WIMP particle annihilating through
three possible final states: $bb$, $W^{+}W^{-}$, and
$\tau^{+}\tau^{-}$.  For the Sculptor dSph, one of the best dSph
candidates in the south, CTA could reach $\sim 10^{-24}$ cm$^2$
s$^{-1}$ at 1~TeV which is comparable to current limits from H.E.S.S.
observations of the GC halo.  For an observation of the GC utilizing
the same 0.3$^\circ$--1.0$^\circ$ annular search region as the
H.E.S.S.  analysis CTA could rule out models with cross sections
significantly below the thermal relic cross section down to $\sim 3
\times 10^{-27}$ cm$^2$ s$^{-1}$.  Overlaid in the figure are WIMP
models generated in the pMSSM framework that satisfy all current
experimental constraints from collider and direct detection searches
\citep{2011EPJC...71.1697C,2012EPJC...72.2156C}.  {\bf Approximately
  half of the models in this set could be excluded at the 95\% C.L. in
  a 500 hour observation of the Galactic center.}

\section{Key Questions and Complementarity with other Techniques}

% For the purpose of community planning for the U.S. HEP to form a
% balanced program for DM, we address several key questions: 
For the purpose of planning for the U.S. HEP community to form a
balanced DM program, we address several key questions: 
(1) {\it
  What are the principal uncertainties in the technique?  What would
  it take to make a convincing detection? }  Gamma-rays provide
excellent calorimetry for almost any annihilation channel, with a
detection cross section that is closely related to the total
annihilation cross section that determines the relic abundance.
Unlike direct detection where the cross-section uncertainties
dominate, astrophysical uncertainties dominate gamma-ray measurements.
For dSphs, these are largely mitigated by the lack of astrophysical
backgrounds, and improving constraints on the halo profile from
dynamical measurements.  However, for the GC, where the prospects for
detection are best, these uncertainties are the largest.  For example,
comparing our baseline NFW MW halo model with the most pessimistic
case of a cored isothermal profile, we calculate a variation of a
factor of $\sim$30 in the J-factor.
% two different halo models (NFW and a cored Isothermal
% profile) we calculate a variation of a factor of $\sim$30 in the
% J-factor. 
The most compelling signature for an actual detection would come from
measuring identical spectra from {\it two different astrophysical
  sources}.  (2) {\it Can one technique like gamma-ray detection
  provide the answer alone, or are other detection methods required?}
Any one method (direct detection, indirect detection with gamma rays,
antimatter or neutrinos) is unlikely to provide a convincing case for
detection.  Each potential signal brings with it the potential for a
new background (a new radioactive decay channel for direct detection
counting experiments, or a new astrophysical source for indirect
detection).  A compelling case probably requires convergent evidence
from a number of different techniques.  But if any accelerator
experiment, or direct detection experiment, were to yield a putative
signal, one clearly would need a gamma-ray experiment to connect the
laboratory discovery to the actual distribution of DM on the sky, and
to help identify the nature of the particle through the details of the
annihilation spectrum.  The ability of all detection techniques to
fully encompass the WIMP parameter space also depends critically on
their complementarity.  For instance CTA will have a unique
sensitivity to very high mass WIMPs (above 1 TeV) that would not be
easily detectable with current accelerator experiments.

\bibliography{cf2_whitepaper_cta_dm}

\begin{thebibliography}{10}
\providecommand*{\bibinfo}[2]{#2}
\providecommand*{\eprint}[1]{#1}
\providecommand*{\url}[1]{#1}
\bibitem{2013arXiv1303.5076P}
\bibinfo{author}{{Planck Collaboration}}, \bibinfo{author}{P.~A.~R. {Ade}},
  \bibinfo{author}{N.~{Aghanim}}, \bibinfo{author}{C.~{Armitage-Caplan}},
  \bibinfo{author}{M.~{Arnaud}}, \bibinfo{author}{M.~{Ashdown}},
  \bibinfo{author}{F.~{Atrio-Barandela}}, \bibinfo{author}{J.~{Aumont}},
  \bibinfo{author}{C.~{Baccigalupi}}, \bibinfo{author}{A.~J. {Banday}},
  \emph{et~al.}, \bibinfo{journal}{ArXiv e-prints}  (\bibinfo{date}{Mar.
  2013}), \eprint{1303.5076}.
\bibitem{2013arXiv1302.6587F}
\bibinfo{author}{J.~L. {Feng}}, \bibinfo{journal}{ArXiv e-prints}
  (\bibinfo{date}{Feb. 2013}), \eprint{1302.6587}.
\bibitem{2011ExA....32..193A}
\bibinfo{author}{M.~{Actis}}, \bibinfo{author}{G.~{Agnetta}},
  \bibinfo{author}{F.~{Aharonian}}, \bibinfo{author}{A.~{Akhperjanian}},
  \bibinfo{author}{J.~{Aleksi{\'c}}}, \bibinfo{author}{E.~{Aliu}},
  \bibinfo{author}{D.~{Allan}}, \bibinfo{author}{I.~{Allekotte}},
  \bibinfo{author}{F.~{Antico}}, \bibinfo{author}{L.~A. {Antonelli}},
  \emph{et~al.}, \bibinfo{journal}{Experimental Astronomy}
  \bibinfo{volume}{\textbf{32}}, \bibinfo{pages}{193} (\bibinfo{date}{Dec.
  2011}), \eprint{1008.3703}.
\bibitem{2004ApJ...616...16G}
\bibinfo{author}{O.~Y. {Gnedin}}, \bibinfo{author}{A.~V. {Kravtsov}},
  \bibinfo{author}{A.~A. {Klypin}}, and \bibinfo{author}{D.~{Nagai}},
  \bibinfo{journal}{\apj} \bibinfo{volume}{\textbf{616}}, \bibinfo{pages}{16}
  (\bibinfo{date}{Nov. 2004}), \eprint{arXiv:astro-ph/0406247}.
\bibitem{2011MNRAS.416.1118C}
\bibinfo{author}{D.~R. {Cole}}, \bibinfo{author}{W.~{Dehnen}}, and
  \bibinfo{author}{M.~I. {Wilkinson}}, \bibinfo{journal}{\mnras}
  \bibinfo{volume}{\textbf{416}}, \bibinfo{pages}{1118} (\bibinfo{date}{Sep.
  2011}), \eprint{1105.4050}.
\bibitem{2012ApJ...744L...9M}
\bibinfo{author}{A.~V. {Macci{\`o}}}, \bibinfo{author}{G.~{Stinson}},
  \bibinfo{author}{C.~B. {Brook}}, \bibinfo{author}{J.~{Wadsley}},
  \bibinfo{author}{H.~M.~P. {Couchman}}, \bibinfo{author}{S.~{Shen}},
  \bibinfo{author}{B.~K. {Gibson}}, and \bibinfo{author}{T.~{Quinn}},
  \bibinfo{journal}{\apjl} \bibinfo{volume}{\textbf{744}}, \bibinfo{pages}{L9},
  \bibinfo{eid}{L9} (\bibinfo{date}{Jan. 2012}), \eprint{1111.5620}.
\bibitem{2008Natur.454..735D}
\bibinfo{author}{J.~{Diemand}}, \bibinfo{author}{M.~{Kuhlen}},
  \bibinfo{author}{P.~{Madau}}, \bibinfo{author}{M.~{Zemp}},
  \bibinfo{author}{B.~{Moore}}, \bibinfo{author}{D.~{Potter}}, and
  \bibinfo{author}{J.~{Stadel}}, \bibinfo{journal}{\nat}
  \bibinfo{volume}{\textbf{454}}, \bibinfo{pages}{735} (\bibinfo{date}{Aug.
  2008}), \eprint{0805.1244}.
\bibitem{2010PhRvD..81d3532K}
\bibinfo{author}{M.~{Kamionkowski}}, \bibinfo{author}{S.~M. {Koushiappas}}, and
  \bibinfo{author}{M.~{Kuhlen}}, \bibinfo{journal}{\prd}
  \bibinfo{volume}{\textbf{81}}(4), \bibinfo{pages}{043532},
  \bibinfo{eid}{043532} (\bibinfo{date}{Feb. 2010}), \eprint{1001.3144}.
\bibitem{2013arXiv1305.0312N}
\bibinfo{author}{D.~{Nieto}}, \bibinfo{author}{M.~{Errando}},
  \bibinfo{author}{L.~{Fortson}}, \bibinfo{author}{B.~{Humensky}},
  \bibinfo{author}{R.~{Mukherjee}}, \bibinfo{author}{M.-{\'A}.
  {S{\'a}nchez-Conde}}, \bibinfo{author}{A.~{Smith}},
  \bibinfo{author}{A.~{Weinstein}}, and \bibinfo{author}{M.~{Wood}},
  \bibinfo{journal}{ArXiv e-prints}  (\bibinfo{date}{May 2013}),
  \eprint{1305.0312}.
\bibitem{2011JCAP...12..011S}
\bibinfo{author}{M.~A. {S{\'a}nchez-Conde}}, \bibinfo{author}{M.~{Cannoni}},
  \bibinfo{author}{F.~{Zandanel}}, \bibinfo{author}{M.~E. {G{\'o}mez}}, and
  \bibinfo{author}{F.~{Prada}}, \bibinfo{journal}{\jcap}
  \bibinfo{volume}{\textbf{12}}, \bibinfo{pages}{11}, \bibinfo{eid}{011}
  (\bibinfo{date}{Dec. 2011}), \eprint{1104.3530}.
\bibitem{2011EPJC...71.1697C}
\bibinfo{author}{J.~A. {Conley}}, \bibinfo{author}{J.~S. {Gainer}},
  \bibinfo{author}{J.~L. {Hewett}}, \bibinfo{author}{M.~P. {Le}}, and
  \bibinfo{author}{T.~G. {Rizzo}}, \bibinfo{journal}{European Physical Journal
  C} \bibinfo{volume}{\textbf{71}}, \bibinfo{pages}{1697} (\bibinfo{date}{Jul.
  2011}), \eprint{1009.2539}.
\bibitem{2012EPJC...72.2156C}
\bibinfo{author}{M.~W. {Cahill-Rowley}}, \bibinfo{author}{J.~L. {Hewett}},
  \bibinfo{author}{S.~{Hoeche}}, \bibinfo{author}{A.~{Ismail}}, and
  \bibinfo{author}{T.~G. {Rizzo}}, \bibinfo{journal}{European Physical Journal
  C} \bibinfo{volume}{\textbf{72}}, \bibinfo{pages}{2156} (\bibinfo{date}{Sep.
  2012}), \eprint{1206.4321}.
\bibitem{2009ApJ...697.1299A}
\bibinfo{author}{E.~{Aliu}}, \bibinfo{author}{H.~{Anderhub}},
  \bibinfo{author}{L.~A. {Antonelli}}, \bibinfo{author}{P.~{Antoranz}},
  \bibinfo{author}{M.~{Backes}}, \bibinfo{author}{C.~{Baixeras}},
  \bibinfo{author}{S.~{Balestra}}, \bibinfo{author}{J.~A. {Barrio}},
  \bibinfo{author}{H.~{Bartko}}, \bibinfo{author}{D.~{Bastieri}},
  \emph{et~al.}, \bibinfo{journal}{\apj} \bibinfo{volume}{\textbf{697}},
  \bibinfo{pages}{1299} (\bibinfo{date}{Jun. 2009}), \eprint{0810.3561}.
\bibitem{2010ApJ...720.1174A}
\bibinfo{author}{V.~A. {Acciari}}, \bibinfo{author}{T.~{Arlen}},
  \bibinfo{author}{T.~{Aune}}, \bibinfo{author}{M.~{Beilicke}},
  \bibinfo{author}{W.~{Benbow}}, \bibinfo{author}{D.~{Boltuch}},
  \bibinfo{author}{S.~M. {Bradbury}}, \bibinfo{author}{J.~H. {Buckley}},
  \bibinfo{author}{V.~{Bugaev}}, \bibinfo{author}{K.~{Byrum}}, \emph{et~al.},
  \bibinfo{journal}{\apj} \bibinfo{volume}{\textbf{720}}, \bibinfo{pages}{1174}
  (\bibinfo{date}{Sep. 2010}), \eprint{1006.5955}.
\bibitem{2011JCAP...06..035A}
\bibinfo{author}{J.~{Aleksi{\'c}}}, \bibinfo{author}{E.~A. {Alvarez}},
  \bibinfo{author}{L.~A. {Antonelli}}, \bibinfo{author}{P.~{Antoranz}},
  \bibinfo{author}{M.~{Asensio}}, \bibinfo{author}{M.~{Backes}},
  \bibinfo{author}{J.~A. {Barrio}}, \bibinfo{author}{D.~{Bastieri}},
  \bibinfo{author}{J.~{Becerra Gonz{\'a}lez}}, \bibinfo{author}{W.~{Bednarek}},
  \emph{et~al.}, \bibinfo{journal}{\jcap} \bibinfo{volume}{\textbf{6}},
  \bibinfo{pages}{35}, \bibinfo{eid}{035} (\bibinfo{date}{Jun. 2011}),
  \eprint{1103.0477}.
\bibitem{2012PhRvD..85f2001A}
\bibinfo{author}{E.~{Aliu}}, \bibinfo{author}{S.~{Archambault}},
  \bibinfo{author}{T.~{Arlen}}, \bibinfo{author}{T.~{Aune}},
  \bibinfo{author}{M.~{Beilicke}}, \bibinfo{author}{W.~{Benbow}},
  \bibinfo{author}{A.~{Bouvier}}, \bibinfo{author}{S.~M. {Bradbury}},
  \bibinfo{author}{J.~H. {Buckley}}, \bibinfo{author}{V.~{Bugaev}},
  \emph{et~al.}, \bibinfo{journal}{\prd} \bibinfo{volume}{\textbf{85}}(6),
  \bibinfo{pages}{062001}, \bibinfo{eid}{062001} (\bibinfo{date}{Mar. 2012}),
  \eprint{1202.2144}.
\bibitem{2008APh....29...55A}
\bibinfo{author}{F.~{Aharonian}}, \bibinfo{author}{A.~G. {Akhperjanian}},
  \bibinfo{author}{A.~R. {Bazer-Bachi}}, \bibinfo{author}{M.~{Beilicke}},
  \bibinfo{author}{W.~{Benbow}}, \bibinfo{author}{D.~{Berge}},
  \bibinfo{author}{K.~{Bernl{\"o}hr}}, \bibinfo{author}{C.~{Boisson}},
  \bibinfo{author}{O.~{Bolz}}, \bibinfo{author}{V.~{Borrel}}, \emph{et~al.},
  \bibinfo{journal}{Astroparticle Physics} \bibinfo{volume}{\textbf{29}},
  \bibinfo{pages}{55} (\bibinfo{date}{Feb. 2008}), \eprint{0711.2369}.
\bibitem{2009ApJ...691..175A}
\bibinfo{author}{F.~{Aharonian}}, \bibinfo{author}{A.~G. {Akhperjanian}},
  \bibinfo{author}{U.~B. {de Almeida}}, \bibinfo{author}{A.~R. {Bazer-Bachi}},
  \bibinfo{author}{B.~{Behera}}, \bibinfo{author}{W.~{Benbow}},
  \bibinfo{author}{K.~{Bernl{\"o}hr}}, \bibinfo{author}{C.~{Boisson}},
  \bibinfo{author}{V.~{Bochow}}, \bibinfo{author}{V.~{Borrel}}, \emph{et~al.},
  \bibinfo{journal}{\apj} \bibinfo{volume}{\textbf{691}}, \bibinfo{pages}{175}
  (\bibinfo{date}{Jan. 2009}), \eprint{0809.3894}.
\bibitem{2011APh....34..608H}
\bibinfo{author}{{H.E.S.S.~Collaboration}}, \bibinfo{author}{A.~{Abramowski}},
  \bibinfo{author}{F.~{Acero}}, \bibinfo{author}{F.~{Aharonian}},
  \bibinfo{author}{A.~G. {Akhperjanian}}, \bibinfo{author}{G.~{Anton}},
  \bibinfo{author}{A.~{Barnacka}}, \bibinfo{author}{U.~{Barres de Almeida}},
  \bibinfo{author}{A.~R. {Bazer-Bachi}}, \bibinfo{author}{Y.~{Becherini}},
  \emph{et~al.}, \bibinfo{journal}{Astroparticle Physics}
  \bibinfo{volume}{\textbf{34}}, \bibinfo{pages}{608} (\bibinfo{date}{Mar.
  2011}), \eprint{1012.5602}.
\bibitem{2011PhRvL.106p1301A}
\bibinfo{author}{A.~{Abramowski}}, \bibinfo{author}{F.~{Acero}},
  \bibinfo{author}{F.~{Aharonian}}, \bibinfo{author}{A.~G. {Akhperjanian}},
  \bibinfo{author}{G.~{Anton}}, \bibinfo{author}{A.~{Barnacka}},
  \bibinfo{author}{U.~{Barres de Almeida}}, \bibinfo{author}{A.~R.
  {Bazer-Bachi}}, \bibinfo{author}{Y.~{Becherini}},
  \bibinfo{author}{J.~{Becker}}, \emph{et~al.}, \bibinfo{journal}{Physical
  Review Letters} \bibinfo{volume}{\textbf{106}}(16), \bibinfo{pages}{161301},
  \bibinfo{eid}{161301} (\bibinfo{date}{Apr. 2011}), \eprint{1103.3266}.
\bibitem{2011PhRvL.107x1302A}
\bibinfo{author}{M.~{Ackermann}}, \bibinfo{author}{M.~{Ajello}},
  \bibinfo{author}{A.~{Albert}}, \bibinfo{author}{W.~B. {Atwood}},
  \bibinfo{author}{L.~{Baldini}}, \bibinfo{author}{J.~{Ballet}},
  \bibinfo{author}{G.~{Barbiellini}}, \bibinfo{author}{D.~{Bastieri}},
  \bibinfo{author}{K.~{Bechtol}}, \bibinfo{author}{R.~{Bellazzini}},
  \emph{et~al.}, \bibinfo{journal}{Physical Review Letters}
  \bibinfo{volume}{\textbf{107}}(24), \bibinfo{pages}{241302},
  \bibinfo{eid}{241302} (\bibinfo{date}{Dec. 2011}), \eprint{1108.3546}.
\bibitem{2012PhRvD..86b2002A}
\bibinfo{author}{M.~{Ackermann}}, \bibinfo{author}{M.~{Ajello}},
  \bibinfo{author}{A.~{Albert}}, \bibinfo{author}{L.~{Baldini}},
  \bibinfo{author}{G.~{Barbiellini}}, \bibinfo{author}{K.~{Bechtol}},
  \bibinfo{author}{R.~{Bellazzini}}, \bibinfo{author}{B.~{Berenji}},
  \bibinfo{author}{R.~D. {Blandford}}, \bibinfo{author}{E.~D. {Bloom}},
  \emph{et~al.}, \bibinfo{journal}{\prd} \bibinfo{volume}{\textbf{86}}(2),
  \bibinfo{pages}{022002}, \bibinfo{eid}{022002} (\bibinfo{date}{Jul. 2012}),
  \eprint{1205.2739}.
\bibitem{2012ApJ...761...91A}
\bibinfo{author}{M.~{Ackermann}}, \bibinfo{author}{M.~{Ajello}},
  \bibinfo{author}{W.~B. {Atwood}}, \bibinfo{author}{L.~{Baldini}},
  \bibinfo{author}{G.~{Barbiellini}}, \bibinfo{author}{D.~{Bastieri}},
  \bibinfo{author}{K.~{Bechtol}}, \bibinfo{author}{R.~{Bellazzini}},
  \bibinfo{author}{R.~D. {Blandford}}, \bibinfo{author}{E.~D. {Bloom}},
  \emph{et~al.}, \bibinfo{journal}{\apj} \bibinfo{volume}{\textbf{761}},
  \bibinfo{pages}{91}, \bibinfo{eid}{91} (\bibinfo{date}{Dec. 2012}),
  \eprint{1205.6474}.
\bibitem{2010JCAP...04..014A}
\bibinfo{author}{A.~A. {Abdo}}, \bibinfo{author}{M.~{Ackermann}},
  \bibinfo{author}{M.~{Ajello}}, \bibinfo{author}{L.~{Baldini}},
  \bibinfo{author}{J.~{Ballet}}, \bibinfo{author}{G.~{Barbiellini}},
  \bibinfo{author}{D.~{Bastieri}}, \bibinfo{author}{K.~{Bechtol}},
  \bibinfo{author}{R.~{Bellazzini}}, \bibinfo{author}{B.~{Berenji}},
  \emph{et~al.}, \bibinfo{journal}{\jcap} \bibinfo{volume}{\textbf{4}},
  \bibinfo{pages}{14}, \bibinfo{eid}{014} (\bibinfo{date}{Apr. 2010}),
  \eprint{1002.4415}.
\bibitem{2013APh....43..189D}
\bibinfo{author}{M.~{Doro}}, \bibinfo{author}{J.~{Conrad}},
  \bibinfo{author}{D.~{Emmanoulopoulos}}, \bibinfo{author}{M.~A.
  {S{\`a}nchez-Conde}}, \bibinfo{author}{J.~A. {Barrio}},
  \bibinfo{author}{E.~{Birsin}}, \bibinfo{author}{J.~{Bolmont}},
  \bibinfo{author}{P.~{Brun}}, \bibinfo{author}{S.~{Colafrancesco}},
  \bibinfo{author}{S.~H. {Connell}}, \emph{et~al.},
  \bibinfo{journal}{Astroparticle Physics} \bibinfo{volume}{\textbf{43}},
  \bibinfo{pages}{189} (\bibinfo{date}{Mar. 2013}), \eprint{1208.5356}.
\bibitem{2012arXiv1210.3503B}
\bibinfo{author}{K.~{Bernl{\"o}hr}}, \bibinfo{author}{A.~{Barnacka}},
  \bibinfo{author}{Y.~{Becherini}}, \bibinfo{author}{O.~{Blanch Bigas}},
  \bibinfo{author}{E.~{Carmona}}, \bibinfo{author}{P.~{Colin}},
  \bibinfo{author}{G.~{Decerprit}}, \bibinfo{author}{F.~{Di Pierro}},
  \bibinfo{author}{F.~{Dubois}}, \bibinfo{author}{C.~{Farnier}}, \emph{et~al.},
  \bibinfo{journal}{ArXiv e-prints}  (\bibinfo{date}{Oct. 2012}),
  \eprint{1210.3503}.
\bibitem{2012AIPC.1505..765J}
\bibinfo{author}{T.~{Jogler}}, \bibinfo{author}{M.~D. {Wood}},
  \bibinfo{author}{J.~{Dumm}}, and \bibinfo{author}{{CTA Consortium}}, in
  \bibinfo{editors}{F.~A. {Aharonian}, W.~{Hofmann}, and F.~M. {Rieger}}, eds.,
  \emph{American Institute of Physics Conference Series} (\bibinfo{date}{Dec.
  2012}), \bibinfo{volume}{vol. 1505 of \emph{American Institute of Physics
  Conference Series}}, \bibinfo{pages}{pp. 765--768}, \eprint{1211.3181}.

\end{thebibliography}

\end{document}